\documentclass[11pt]{article}
\usepackage{graphicx}

\usepackage{graphicx}

\newcommand{\BABARConfYear}    {02}

\newcommand{\BABARConfNumber}  {034}

\newcommand{\SLACPubNumber} {9302}
\newcommand{\LANLNumber} {0207053}
\input pubboard/babarsym

\def\bdspi    {\ensuremath {\Bz{\to}D_s^{+}\pi^-}}
\def\bdsspi    {\ensuremath {\Bz{\to}D_s^{*+}\pi^-}}
\def\bdsk    {\ensuremath {\Bz{\to}D_s^- K^+}}
\def\bdssk    {\ensuremath {\Bz{\to}D_s^{*-}K^+}}
\def\bdsordsspi   {\ensuremath {\Bz{\to}D_s^{(*)+}\pi^-}}
\def\bdsordssk   {\ensuremath {\Bz{\to}D_s^{(*)-}K^+}}

\def\sin2bg       {\ensuremath {\sin (2\beta+\gamma)}}
\def\btou       {\ensuremath {b\rightarrow u}}

\def\brdsShort {\ensuremath { 3.2 \pm 0.9\, ({\rm stat. }) }}

\def\brdskShort {\ensuremath{ 3.2 \pm 1.0\, ({\rm stat. }) }}

\def\brds {\BR(\bdspi)~=~(\brdsShort $\pm 1.0\, ({ \rm syst. })$) \ensuremath{\times 10^{-5}}}

\def\brdsk {\BR(\bdsk)~=~(\brdskShort$\pm 1.0\, ({ \rm syst. })$) \ensuremath{\times 10^{-5}}}

\def\brdsslim {\BR(\bdsspi)\ensuremath{<4.1{\times}10^{-5}}}
\def\brdssklim {\BR(\bdssk)\ensuremath{<2.5{\times}10^{-5}}}

\def\lumi    {\ensuremath {84\ \rm million}}
\def\mds     {\ensuremath {M_{D_s}^{cand}}}

\def\cth     {\ensuremath {\cos{\theta_{T}}}}
\def\cthel     {\ensuremath {\cos{\theta_{H}}}}
\def\ctwothel     {\ensuremath {\cos^2{\theta_{H}}}}

\def\de        {\ensuremath {\Delta E}}

\def\myprl  #1 #2 #3 {\jprl{#1},\ #2 (#3)}
\def\myplb  #1 #2 #3 {\plb{#1},\ #2 (#3)}
\def\myprd  #1 #2 #3 {\jprd{#1},\ #2 (#3)}
\def\mynim  #1 #2 #3 {\nim{#1},\ #2 (#3)}
\def\mypr   #1 #2 #3 {\pr{#1},\ #2 (#3)}

\def\myzpc  #1 #2 #3 {\zpc{#1},\ #2 (#3)}

\long\def\inst#1{\par\nobreak\kern 4pt\nobreak
    {\it #1}\par\vskip 10pt plus 3pt minus 3pt}

\begin{document}
{\pagestyle{empty}

\begin{flushright}
\babar-CONF-\BABARConfYear/\BABARConfNumber\\
SLAC-PUB-\SLACPubNumber\\
hep-ex/\LANLNumber\\
July, 2002 \\
\end{flushright}

\par\vskip 5cm

\begin{center}
\Large \bf A Study of the Rare Decays   \bdsordsspi\ and \bdsordssk
\end{center}

\bigskip

\begin{center}
\large The \babar\ Collaboration\\
\mbox{ }\\
\today
\end{center}
\bigskip \bigskip

\begin{center}
\large \bf Abstract
\end{center}
We report  on the evidence for  the  decays \bdspi\  and \bdsk\ 
 and the results of a search for the decays \bdsspi\ and \bdssk\
from a sample  of \lumi\ \FourS\ decays into $B$ meson pairs collected with the \babar\ detector
at the PEP II asymmetric-energy $e^{+}e^{-}$ collider. 
 The measured \bdspi\ yield has a probability of less than $10^{-3}$ to be a
 fluctuation of the background 
and we measure the branching fraction \brds. The measured \bdsk\ yield has a probability of less than 
$5 \times 10^{-4}$ to be a fluctuation of the background 
and we measure the branching fraction \brdsk.  We also set 90\% C.L. limits
 \brdsslim\ and \brdssklim. All results are preliminary.

\vfill
\begin{center}
Contributed to the 31$^{st}$ International Conference on High Energy Physics,\\ 
7/24---7/31/2002, Amsterdam, The Netherlands
\end{center}

\vspace{1.0cm}
\begin{center}
{\em Stanford Linear Accelerator Center, Stanford University, 
Stanford, CA 94309} \\ \vspace{0.1cm}\hrule\vspace{0.1cm}
Work supported in part by Department of Energy contract DE-AC03-76SF00515.
\end{center}

\newpage
} 

\begin{center}
\small

The \babar\ Collaboration,
\bigskip

B.~Aubert,
D.~Boutigny,
J.-M.~Gaillard,
A.~Hicheur,
Y.~Karyotakis,
J.~P.~Lees,
P.~Robbe,
V.~Tisserand,
A.~Zghiche
\inst{Laboratoire de Physique des Particules, F-74941 Annecy-le-Vieux, France }
A.~Palano,
A.~Pompili
\inst{Universit\`a di Bari, Dipartimento di Fisica and INFN, I-70126 Bari, Italy }
% ...

\end{center}\newpage

\newpage

The measurement of the \CP-violating phase of the Cabibbo-Kobayashi-Maskawa  (CKM) matrix~\cite{CKM}
is an important part of the present scientific program in particle physics. 
\CP\ violation manifests itself as a non-zero area of the unitarity triangle~\cite{Jarlskog}.
While it is sufficient to measure one of the angles to demonstrate the
existence of \CP\ violation,
the unitarity triangle needs to be overconstrained by experimental measurements,
in order to demonstrate that the CKM mechanism is the correct 
explanation of this phenomenon. Several theoretically clean measurements
of the angle $\beta$ exist~\cite{sin2b}, but there is no such
measurement of the two other angles $\alpha$ and 
$\gamma $. A theoretically clean measurement of $\rm sin(2\beta+\gamma)$ can be obtained from the study
of the time evolution of the  $\Bz{\to} D^{(*)-} \pi^+$~\cite{chconj} decays, of which a large sample
is already available at the B-factories, and of the corresponding
Cabibbo suppressed mode $\Bz{\to} D^{(*)+}\pi^-$~\cite{sin2bg}.
 This measurement 
requires the knowledge of the ratio 
between the decay amplitudes $R_{\lambda}^{(*)}=|A(\Bz{\to} D^{(*)+}\pi^-)/A(\Bz{\to}
D^{(*)-}\pi^+)|$. 
Unfortunately the measurement of $|A(\Bz{\to} D^{(*)+}\pi^-)|$ via the
measurement of $\BR(\Bz{\to} D^{(*)+}\pi^-)$ is not
possible with the currently available data sample due to the presence
of the copious background from $\Bzb{\to} D^{(*)+}\pi^-$.
However, we can  measure \BR($\bdsordsspi$) and
relate it to $R_{\lambda}^{(*)}$  using SU(3) symmetry:
$R_{\lambda}^{(*)2}\propto \frac{\BR(\bdsordsspi)}{\BR(\Bzb{\to} D^{(*)-}\pi^+)} $,
where the proportionality constant is, to first approximation, the ratio
of the $D_s^{(*)+}$ and the $D^{(*)+}$ decay constants~\cite{sin2bg}.
The decays \bdsordsspi\ have also been proposed to be used for the
measurement of   $ |V_{ub}/V_{cb}|$~\cite{roy}.

The decays \bdsordssk\ are a probe of the dynamics in $B$ decays because they
are expected to proceed  mainly via a
W-exchange diagram, not observed so far. In addition, theses modes  can
be used to investigate the role of final state  rescattering since its 
presence can substantially  increase the expected rates~\cite{Wexch}.
In this letter we present  measurements of the
branching fractions for the decays  \bdsordsspi  and \bdsordssk.

This analysis uses  a sample  of \lumi\ \FourS\ decays into $B\overline{B}$
 pairs 
 collected  in the years 1999-2002 with the \babar\ detector
at the \pep2\ asymmetric-energy $B$-factory~\cite{pep}.
Since the \babar\ detector is described in detail
elsewhere~\cite{detector},  
only the components of the detector crucial to this analysis are
 summarized below. 
Charged particle tracking is provided by a five-layer silicon
vertex tracker (SVT) and a 40-layer drift chamber (DCH). 
For charged-particle identification, ionization energy loss ($dE/dx$) in
 the DCH and SVT, and Cherenkov radiation detected in a ring-imaging
 device are used. 
Photons are identified and measured using
the electromagnetic calorimeter, which comprises 6580 thallium-doped CsI
crystals. These systems are mounted inside a 1.5 T solenoidal
superconducting magnet. 
We use the GEANT~\cite{geant} software to simulate interactions of particles
traversing the \babar\ detector, taking into account the varying detector conditions and beam backgrounds.

We select events with a minimum of three reconstructed charged tracks and
a total measured energy greater than 4.5 GeV as determined
using all charged tracks and neutral clusters with energy above 30 MeV.
In order to reject continuum background, the ratio of the second
and zeroth order Fox-Wolfram moments~\cite{fox} 
must be less than 0.5.

So far, only upper limits on the modes studied in this letter
exist~\cite{prior}. Therefore 
the selection criteria are optimized
 to maximize the ratio of signal efficiency over the square-root of the
expected number of background events.

The \Ds\ mesons 
are reconstructed in the modes
 $\Ds {\to} \phi \pi^+$, $\KS K^+$ and $\Kstarzb\Kp$, 
with $\phi{\to} K^+K^-$, $\KS {\to} \pip \pim$ and $\Kstarzb{\to} K^-\pi^+$. 
 The $\KS$ candidates are reconstructed from two
oppositely-charged tracks with an invariant mass 
$493 < M_{\pip \pim} < 501\mevcc $. All other tracks are 
required  to originate from a vertex consistent with the
$\epem$ interaction point.
In order to identify charged kaons, two selections are used: a pion
veto with an efficiency of 95\% for kaons and a 20\% 
pion misidentification, and a tight kaon selection 
with an efficiency of 85\%  and 5\% pion misidentification probability.
If not otherwise specified, the pion veto is always adopted. 
The $\phi$ candidates are reconstructed from two oppositely-charged 
kaons with an invariant mass $1009 < M_{\Kp \Km} < 1029\mevcc $. 
The $\Kstarzb$ candidates are constructed from the  \Km\ and a \pip\
candidates and are required to have an invariant mass in the range
$856 < M_{\Km \pip} < 936\mevcc $.
The polarizations of the 
 \Kstarzb\ ($\phi$) mesons in the   
\Ds\ decays are also utilized to reject backgrounds through the use of the helicity angle
$\theta_H$, defined as the angle between one of the decay products  of
the \Kstarzb\ ($\phi$) 
and the direction of flight of the meson itself, in the meson rest
frame. 
Background events are distributed uniformly in $\cthel$ since they
originate from random combinations, while
signal events are distributed as $\ctwothel$.
The \Kstarzb\ candidates are therefore required to have $|\cthel|>0.4$, while
for the $\phi$ candidates we require $|\cthel|>0.5$. 
In order to reject background from \Dp{\to}\KS\pip or \Kstarzb\pip,
the $\Kp$ in the reconstruction of \Ds{\to}\KS\Kp or \Kstarzb\Kp is
required to pass the tight 
kaon identification criteria introduced above.
Finally, the \Ds\ candidates are required to have an invariant mass within 10 \mevcc\ 
of the nominal mass~\cite{PDG2002}.

We reconstruct  \Dss\ candidates in the mode  $\Dss{\to}\Ds\gamma$, by combining \Ds\ and photon candidates. 
Photons that form a $\pi^0$ candidate, with $122< M_{\gamma\gamma}<147
 \mevcc$,
 in combination  with any other photon  with energy greater than 70 \mev are rejected.
 The mass difference between the \Dss\ and  the \Ds\ candidate is
 required to be within 14 \mevcc\ of 
the nominal value~\cite{PDG2002}. 

\begin{figure}[!htb]
\begin{center}

\includegraphics[width=0.6\linewidth]{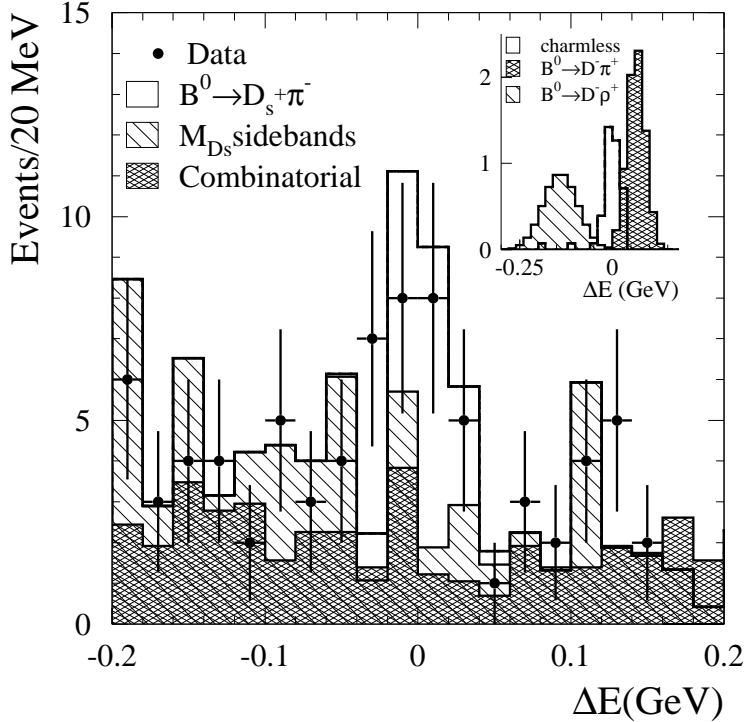}
\label{fig:datade}
\caption{The \de\ distribution in data compared with the distribution
  in the combinatorial background, estimated from the \mes\ sidebands,
  and with the cross-contamination, which is estimated from the \mds\
  sidebands.
The insert shows separately the  \de\ distribution of the
  contributions to the cross contamination as expected from the
  simulation. The reflection background is normalized to the known
  branching fractions~\cite{PDG2002}, while the normalization of the
  charmless background
is arbitrary.
}
\end{center}
\end{figure}

We combine  \Ds\ or \Dss\ candidates with a track of opposite
charge to form $\bdsordssk$  or $\bdsordsspi$  candidates
depending on whether they pass the tight kaon selection criteria. 
In order to reject events where the \Ds
comes from a $B$ candidate and the pion or kaon from the other $B$, we require the two candidates to
have a probability greater than 0.25\% of originating from a common
vertex. 
The remaining background is predominantly combinatorial in nature and
arises from continuum \qqbar\
production. 
In order to suppress it using the event topology, we compute the angle
($\theta_T$) between the thrust axis of 
 the  $B$ meson decay product candidates and the thrust axis of  all the other particles in the event.
In the center-of-mass frame (c.m.), \BB\ pairs are
produced approximately at rest and  produce a uniform $\cth$ distribution.
In contrast,
\qqbar\ pairs are produced back-to-back in the c.m. frame,
which results in a $|\cth|$
 distribution peaking at 1. 
Depending on the background level of each mode, $|\cth|$ is required to
be smaller than a value which ranges between 0.7 and 0.8.
 We further suppress backgrounds using a Fisher discriminant $\cal{F}$ constructed from
the scalar sum of the c.m. momenta of all tracks and photons (excluding the $B$ candidate
decay products)  flowing into 9 concentric cones centered on the thrust axis of the $B$ candidate~\cite{twobody}. The more spherical the
event, the lower the value of ${\cal{F}}$. 
We require ${\cal{F}}$ to be smaller than a threshold which varies from
  0.04 to 0.2 depending on the background level.
 
We extract the signal using the kinematic variables
$\mes= \sqrt{E_{\rm b}^{*2} - (\sum_i {\bf p}^*_i)^2}$
and $\de= \sum_i\sqrt{m_i^2 + {\bf p}_i^{*2}}- E_{\rm b}^*$,
where $E_{\rm b}^*$ is the beam energy in the c.m. frame,
${\bf p}_i^*$ is the c.m.~momentum of daughter particle $i$ of the
$B$ meson candidate, and $m_i$ is the mass hypothesis for particle $i$. 
For signal events, $m_{\rm ES}$ peaks at the $B$ meson mass with 
a resolution of about 2.5 MeV$/c^2$ and $\Delta E$ peaks near zero,  
indicating that the candidate system of particles has total energy consistent with
the beam energy in the c.m.~frame.
The 
  \de\ signal band is defined by $|\de|<36 \mev$ and within it
we define as signal candidates the events with $\mes>5.27$\gevcc.

After the aforementioned selection, three classes of backgrounds remain.
First, the amount of {\it{combinatorial background}} in the signal region is 
estimated from the sidebands of 
the \mes\ distribution and  is
described by a threshold function 
$\frac{dN}{dx}=x\sqrt{1-x^{2}/E_b^{*2}}\exp\left[-\xi\left(1-x^{2}/E_b^{*2}\right)\right]$, 
characterized by the shape parameter $\xi$~\cite{argus}. 

\begin{figure}[!htb]
\begin{center}

\includegraphics[width=\linewidth]{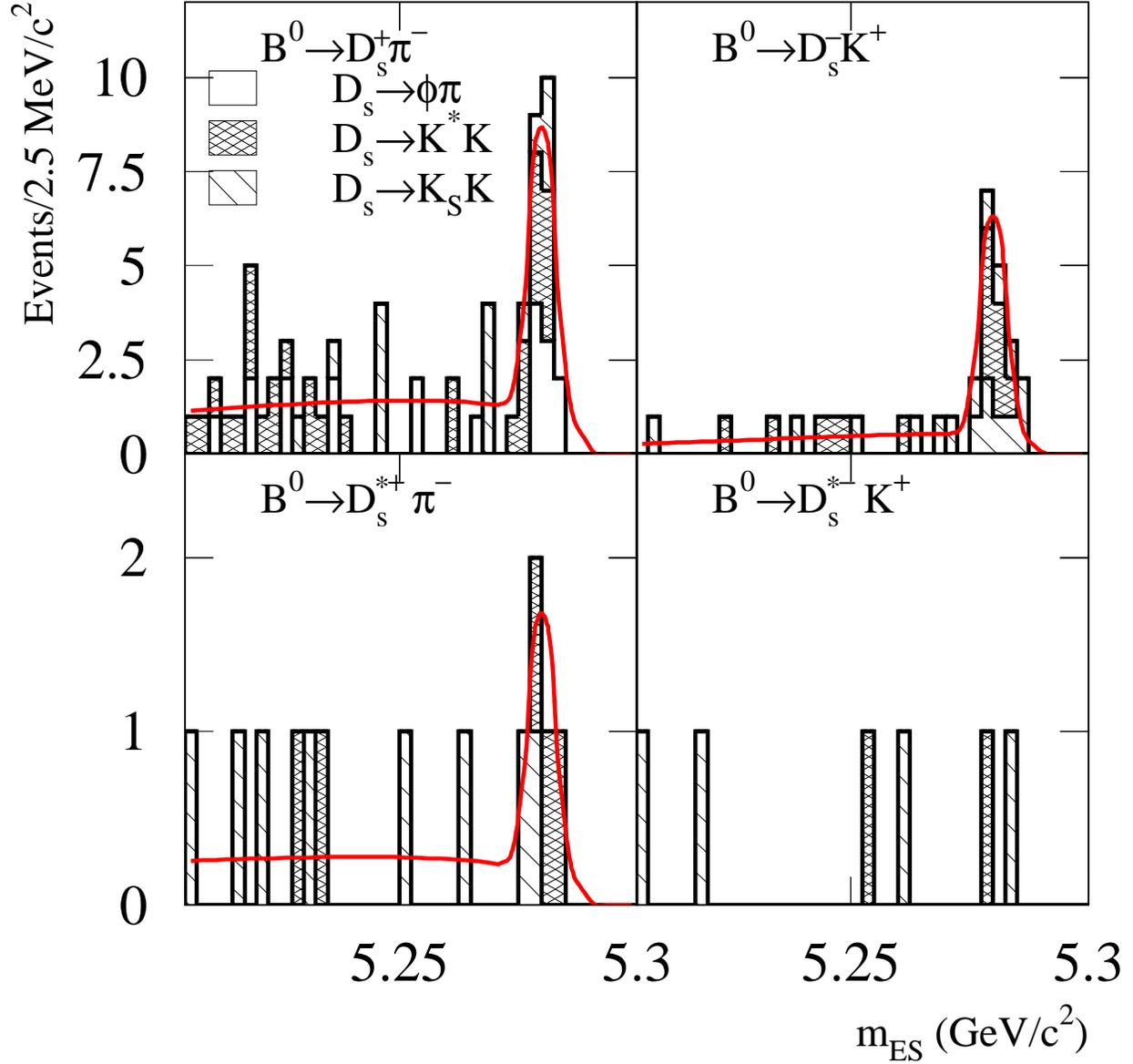}
\label{fig:datamb}

\caption{The \mes\ distributions for the  \bdspi\ (top left), \bdsspi\
  (top right), \bdsk\ (bottom left), and \bdssk\
  (bottom right)
candidates on data  after the selection, within the \de\ band. 
The fits used to obtain the signal yield are described in the text. The
  contribution from 
each \Ds\ mode is shown. 
}
\end{center}
\end{figure}

Next,  $B$ meson decays such as  $\Bzb{\to}\Dp\pim,\rho^-$ with
$\Dp{\to}\KS\pip$ or $\Kstarzb\pip$ can 
constitute a background for  the \bdspi\ mode if
 the pion in the $D$ decay is misidentified as a kaon ({\it{reflection
background}}). This background has the same \mes\ distributions 
as the signal but different distributions of \de.
The corresponding background for the
\bdsk\ mode ($\Bz{\to}\Dm\Kp,\Kstarp$) has a branching fraction ten times smaller.
Finally, rare $B$  decays into the same final state, such as
$\Bz{\to} \Kbar^{(*)0}\Kp\pim$ or $\Kbar^{(*)0}\Kp\Km$
({\it{charmless background}}),
have the same \mes\ and \de\ distributions as the signal.
Figure \ref{fig:datade} shows the \de\ distribution for the signal and
for the various sources of background.

The branching fraction of the charmless background is not well measured
and we therefore need to  estimate the sum of the reflection and charmless background (referred
to as {\it{cross-contamination}}) directly on data. This is possible
because both of these backgrounds have a flat distribution in
the \Ds\ candidate (\mds) mass while the signal has a Gaussian distribution.
Possible peaking background 
from $B{\to}\Ds X$ decays is negligible, as determined from simulation. 
The cross-contamination to the decays \bdsspi\ and \bdssk\ is dominated
by the reflection background which we  estimate from simulation.
Cross-feed between \bdsordsspi\ and \bdsordssk\ modes has been estimated
 to be less than 1\%.


Figure~\ref{fig:datamb} shows the \mes\ distribution for each of the
modes in the \de\ signal band. 
We perform an unbinned
maximum-likelihood fit  to each
 \mes\ distribution with the threshold function to characterize the
 combinatorial background and a Gaussian function to describe the sum of 
the signal and cross-contamination contributions. 
 The mean and the width of the Gaussian distribution are fixed to the 
values obtained in a copious $\Bz{\to} D^{(*)-} \pip$ control sample.
 For the \bdspi\ and \bdsk\ analyses, we obtain the threshold parameter $\xi$  from a
 fit to the data distributions of \mes\ after loosening the \mds\ and
 \de\ requirements. In the case of \bdsspi\ and \bdssk, due to the low 
 background level,  we use simulated events to estimate $\xi$.

No fit is performed to the \bdssk sample, due to the low number of events.
Whenever there are enough events we perform a fit to each \Ds\ decay
mode separately, as well as on the combination of all modes.
The cross-contamination is estimated performing the same fit on the
events in the data \mds\ sidebands ($4\sigma<|\mds-1968.6\mevcc|<8\sigma$,
where the resolution is $\sigma_{\mds}=5\mevcc$).
The number of observed events, the background expectations and the 
reconstruction efficiencies as estimated on simulated events are
summarized in Table~\ref{tab:fit}.


\begin{table*}[t!]
\caption{The number of signal candidates ($N_{sigbox}$ ), 
the  Gaussian yield ($N_{gaus}$) and  the combinatorial background ($N_{comb}$) as extracted from the likelihood fit, 
the reconstruction efficiency ($\varepsilon$), the cross-contamination ($N_{cross}$),
 the probability ($P_{bckg}$) of the data being consistent with 
the background fluctuating up to the level of the data in the 
absence of signal, the measured branching fraction (\BR
), and the 90\% confidence level upper limit. 
$N_{gaus}$, $N_{comb}$ and \BR\  are not available for modes with too few events.
$N_{cross}$ is not reported if no event is found in the \Ds\ mass sideband.
}

\begin{center}
\mbox{ \scriptsize
\begin{tabular}{l|c|c|c|c|c||c|c|c} \hline \hline
$B$ mode& $N_{sigbox}$ & $N_{gaus}$  & $N_{comb}$  & $N_{cross}$ &
$\varepsilon$(\%)&  $P_{bckg}$ & \BR ($10^{-5}$)  & 90\%
C.L. \\
& & & & & & & & ($10^{-5}$)\\ \hline
\bdspi & & & & & & & & \\
\ \ $\Ds{\to}\phi\pip$    &9 &$8.0\pm3.0$  &$2.1\pm0.7$ &$<0.7$ &16.9  &$1.4\times10^{-3}$ & $3.1 \pm 1.2$ & -\\
\ \ $\Ds{\to}\Kstarzb\Kp$ &12 &$9.2\pm3.4$ &$3.8\pm1.0$ &$2.9\pm1.8$ & 9.6  &$2.3\times10^{-2}$ & $3.5 \pm 1.9$ & -\\
\ \ $\Ds{\to}\KS\Kp$      &5 &$4.2\pm2.2$  &$1.9\pm0.6$ &$1.2\pm1.4$ &12.3  &$8.3\times10^{-2}$ &$2.4 \pm 1.8$  & -\\ 
\ \ all                   &26 &$21.4\pm 5.1$ &$7.8\pm1.7$ &$3.7\pm2.4$       &N/A   &$9.5\times10^{-4}$ & $3.2\pm 0.9\pm 1.0$&-\\ \hline
\bdsspi & & & & & & & & \\
\ \ $\Ds{\to}\phi\pip$    &2 &- &$0.6\pm0.3$ &$<0.14$ &7.8  &- &- &- \\
\ \ $\Ds{\to}\Kstarzb\Kp$ &3 &$2.8_{-1.8}^{+2.7}$ & $0.4\pm0.3$ & $0.3\pm0.2$ & 3.3  &$3.9\times 10^{-2}$  &$4.3_{-3.1}^{+4.7}$  &$<12$  \\
\ \ $\Ds{\to}\KS\Kp$      &0 &- &$0.4\pm0.3$ & $<0.14$ & 5.1  &- &-&-\\ 
\ \ all                   &5 &$4.4_{-2.8}^{+2.7}$ &$1.2\pm0.4$  &$0.3\pm0.2$ &N/A & $2.3\times 10^{-2}$ & $1.9_{-1.3}^{+1.2}\pm0.5$  & $<4.1$ \\ \hline
\bdsk  & & & & & & & & \\
\ \ $\Ds{\to}\phi\pip$    &7 &$5.8\pm2.6$ &$1.3\pm0.7$ &$1.1\pm1.2$ &13.0  &$4.5\times10^{-2}$ &$2.4\pm1.3$ &- \\
\ \ $\Ds{\to}\Kstarzb\Kp$ &8 &$7.3\pm2.9$ &$1.7\pm0.7$ &$<0.7$ &7.8  &$1.9\times10^{-3}$ &$5.0\pm2.0$  &- \\
\ \ $\Ds{\to}\KS\Kp$      &4 &$3.7\pm2.0$ &$0.6\pm0.4$ &$1.3\pm1.0$ &9.2  &$1.7\times10^{-2}$ &$2.5\pm2.1$ &- \\ 
\ \ all                   &19 &$16.7\pm4.3$ &$3.5\pm1.3$ &$2.7\pm1.9$ &N/A &$5.0\times10^{-4}$  & $3.2\pm1.0\pm1.0$ & -\\ \hline
\bdssk & & & & & & & & \\
\ \ $\Ds{\to}\phi\pip$    &0 &- &$0.8\pm0.6$ &$<0.14$&5.3 &- &- &- \\
\ \ $\Ds{\to}\Kstarzb\Kp$ &1 &- &$0.4\pm0.4$ &$<0.14$ &2.7 &- &- &- \\
\ \ $\Ds{\to}\KS\Kp$      &1 &- &$0.4\pm0.4$ &$<0.14$  &4.3 &- &- &- \\ 
\ \ all                   &2 &- &$1.6\pm0.8$ &$<0.14$  &N/A &0.48 &- &$<2.5$ \\ \hline
\end{tabular}}
\end{center}
\label{tab:fit}
\end{table*}

In the \bdspi (\bdsk) mode the fit yields a
Gaussian contribution of $21.4\pm 5.1$ ($16.7\pm 4.3$) events and a
combinatorial background of $7.8\pm 1.7$ ($3.5\pm 1.3$) events. The
cross-contamination is 
estimated to be $3.7\pm 2.4$ ($2.7\pm 1.9$) events. The probability of
the background to fluctuate to the observed number of events, taking
into account both Poisson fluctuations and uncertainties in the
background estimates, is $9.5\times 10^{-4}$ ($5.0\times
10^{-4}$). For a Gaussian distribution this would correspond to
$3.3\sigma$ ($3.5\sigma$).
Given the estimated reconstruction efficiencies we measure
$\BR(\bdspi)=(3.2\pm 0.9)\times 10^{-5}$ ($\BR(\bdsk)=(3.2\pm 1.0)\times 10^{-5}$),
where the quoted error is statistical only.
We also set the  90\% C.L. limits  \brdsslim\ and \brdssklim.
All results are preliminary.

The systematic errors are dominated by the 25\% relative uncertainty in  
 \BR (\Ds$\rightarrow\phi\pip$). The uncertainties on the knowledge of
 the background come from uncertainties in the $\xi$ parameter, for the
 combinatorial background, and from  the limited number of events in the
 \mds\ sidebands for the cross-contamination. They amount to 14\%, 16\%, 7\% and 36\% of the measured
 branching fractions in the  \bdspi , \bdsk, \bdsspi and \bdssk\ modes respectively.
The rest of the systematic errors, which include the uncertainty on tracking, \KS\ and charged kaons identification
 efficiencies range between   11\% and 14\% depending on the mode.

 
In conclusion, 
we  report a 3.3$\sigma$ signal for the \btou\ transition 
 \bdspi and a 3.5$\sigma$ signal for the \bdsk\ decay, and we determine
the preliminary results
\begin{center}
 \brds \ \\
 \brdsk.
\end{center}
Since the dominant uncertainty 
comes from the knowledge of the \Ds\ branching fractions we also compute
$\BR( \bdspi )\times \BR(\Ds\rightarrow\phi\pip)=(1.13\pm0.33\pm0.21)
\times 10^{-6}$
and $\BR( \bdsk )\times \BR(\Ds\rightarrow\phi\pip)=(1.16\pm0.36\pm0.24)
\times 10^{-6}$.
The search for \bdsspi\ and \bdssk\  yields the preliminary 90\% C.L. upper limits
\begin{center}
 \brdsslim \  \\
 \brdssklim.
\end{center}

We are grateful for the 
extraordinary contributions of our \pep2\ colleagues in
achieving the excellent luminosity and machine conditions
that have made this work possible.
The success of this project also relies critically on the 
expertise and dedication of the computing organizations that 
support \babar.
The collaborating institutions wish to thank 
SLAC for its support and the kind hospitality extended to them. 
This work is supported by the
US Department of Energy
and National Science Foundation, the
Natural Sciences and Engineering Research Council (Canada),
Institute of High Energy Physics (China), the
Commissariat \`a l'Energie Atomique and
Institut National de Physique Nucl\'eaire et de Physique des Particules
(France), the
Bundesministerium f\"ur Bildung und Forschung and
Deutsche Forschungsgemeinschaft
(Germany), the
Istituto Nazionale di Fisica Nucleare (Italy),
the Research Council of Norway, the
Ministry of Science and Technology of the Russian Federation, and the
Particle Physics and Astronomy Research Council (United Kingdom). 
Individuals have received support from 
the A. P. Sloan Foundation, 
the Research Corporation,
and the Alexander von Humboldt Foundation.


\begin{thebibliography}{99}
\bibitem{chconj}
Charge conjugation is implied throughout this letter, unless 
explicitely stated.
\end{thebibliography}
\end{document}